\documentclass[aps,prl,twocolumn,superscriptaddress,nofootinbib,floatfix]{revtex4-2}
\usepackage{amsmath,amssymb,bm,booktabs}
\usepackage[colorlinks=true,linkcolor=blue,citecolor=blue,urlcolor=blue]{hyperref}
\usepackage{amsmath,amssymb,bm,booktabs,graphicx}

\newcommand{\tr}{\operatorname{tr}}
\newcommand{\EH}{\mathrm{EH}}
\newcommand{\RP}{\mathbb{RP}^3}
\newcommand{\CS}{\mathrm{CS}}
\newcommand{\Hil}{\mathcal H}
\newcommand{\cO}{\mathcal O}

\makeatletter
\def\paragraph{\@startsection{paragraph}{4}{\parindent}{\z@}{-1em}{\normalfont\normalsize\bfseries}}
\makeatother

\begin{document}

\title{Center anomalies without twists:\\ torsion, exact degeneracies, and condensates}

\author{Mohamed M. Anber}
\affiliation{Centre for Particle Theory, Department of Mathematical Sciences, Durham University, South Road, Durham DH1 3LE, UK}

\begin{abstract}
Mixed center anomalies are usually exposed by 't~Hooft's twisted boundary conditions in a Hamiltonian formalism. On $\RP$, torsion makes the twist unnecessary: a center operator on the untwisted $SU(2)$ Hilbert space anticommutes with time reversal in Yang--Mills theory at $\theta=\pi$ and with the chiral symmetry of super Yang--Mills theory, so every level is at least doubly degenerate. The relevant semiclassical object is the BPST instanton, which descends to a charge-$\tfrac12$ saddle on the orbifold $\mathbb R^4/\mathbb Z_2$.  We complete it in two ways: applying a conformal transformation maps the orbifold to $\mathbb R\times\RP$, with the saddle describing the tunneling event between the two center sectors; by resolving the orbifold singularity, we obtain the Eguchi--Hanson space, which gives the gaugino condensate at its weak-coupling value.
\end{abstract}

\maketitle

\paragraph{Introduction.}
Generalized global symmetries~\cite{Gaiotto:2014kfa} and their 't~Hooft anomalies constrain the phases of gauge theories. In 4D $SU(2)$ Yang--Mills (YM) theory at $\theta=\pi$, a mixed anomaly between time reversal and the $\mathbb Z_2$ center ($1$-form) symmetry forbids a trivially gapped vacuum~\cite{Gaiotto:2017yup}. In $\mathcal N=1$ super Yang--Mills (SYM) theory, a mixed anomaly between the discrete chiral symmetry $\mathbb Z_{4}^\chi$ and the center forces a chiral condensate \cite{Shimizu:2017asf,Komargodski:2017smk}. In Euclidean language, these anomalies require a background two-form gauge field. In Hamiltonian language, they require 't~Hooft's twisted boundary conditions on $\mathbb T^3$~\cite{tHooft:1979rtg,vanBaal:1982ag,vanBaal:thesis,Gonzalez-Arroyo:1997ugn,vanBaal:2000zc,Cox:2021vsa,Anber:2023mlc}, i.e., a fractional magnetic flux, as was first pointed out in \cite{Cox:2021vsa}: without it, center-related flat connections on $\mathbb T^3$ have Chern--Simons (CS) invariants that differ only by integers, and the anomaly is invisible.

\begin{figure}[t]
\includegraphics[width=\columnwidth]{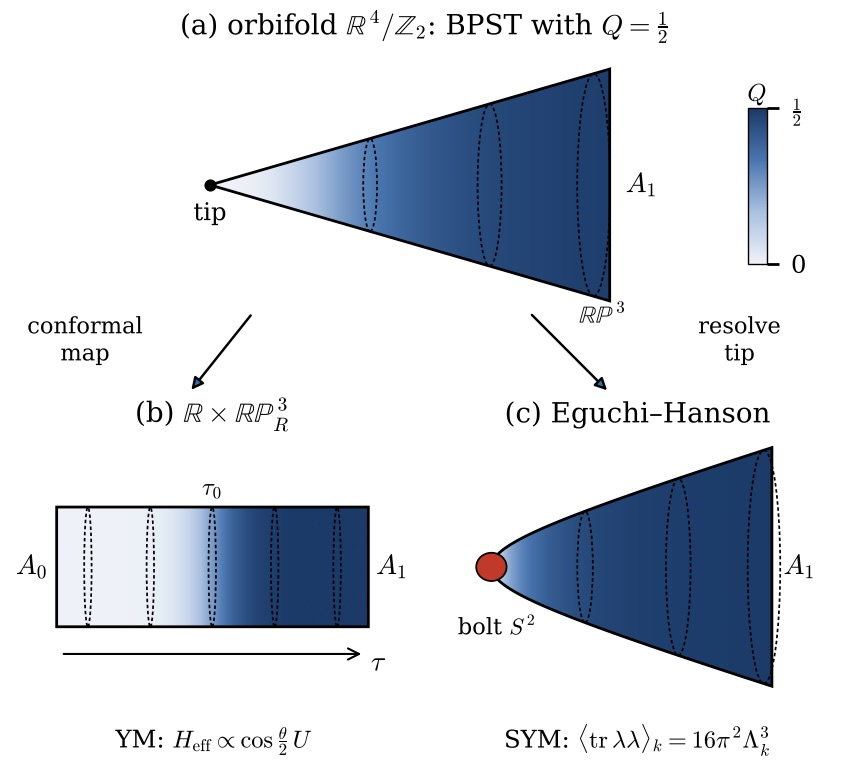}
\caption{(a) The BPST instanton centered at the origin descends to a $Q=\frac12$ configuration on the orbifold $\mathbb R^4/\mathbb Z_2$, with asymptotic flat connection $A_1$ on $\RP$. (b) The conformal map sends the tip to $\tau\to-\infty$: on $\mathbb R\times\RP_R$ it is a tunneling event from $A_0$ to $A_1$. (c) Resolving the tip gives EH space, a cap whose bolt carries $\CS=0$. Shading: charge enclosed, from $0$ to $\frac12$.}
\label{fig:completions}
\end{figure}

In this Letter, we point out that for $SU(2)$ (S)YM theory on $\RP=S^3/\mathbb Z_2$ no twist is needed. The torsion of $\RP$ does the job of the magnetic flux. The untwisted $SU(2)$ Hilbert space contains two flat sectors whose CS invariants differ by $\tfrac12$, together with an operator $U$ that implements the center symmetry and exchanges them. We derive the following results.
\begin{itemize}
\item[(a)] In SYM theory, and in YM theory at $\theta=\pi$, $U$ anticommutes with the chiral generator and with time reversal, respectively. Every energy level is at least doubly degenerate at any radius of $\RP$. The twisted partition function $\mathrm{Tr}[Ue^{-\beta H}]$, proposed recently as a diagnostic of monopole condensation~\cite{Hayashi:2026brp}, vanishes identically in both theories. 
\item[(b)] One semiclassical object controls the dynamics: the BPST instanton centered at the origin descends to a charge-$\tfrac12$ self-dual configuration on the orbifold $\mathbb R^4/\mathbb Z_2$ that connects the two sectors. Its tip can be completed in two ways (Fig.~\ref{fig:completions}). A conformal map gives the cylinder $\mathbb R\times\RP$, where the configuration is a tunneling event that yields $\mathrm{Tr}[Ue^{-\beta H}]$ in closed form at small radius in YM theory. Resolving the tip gives the Eguchi--Hanson (EH) space~\cite{Eguchi:1978xp}, on which the same saddle gives the SYM gaugino condensate at its weak-coupling value.
\item[(c)] For $SU(N)$ on the lens spaces $L(p,1)\equiv S^3/\mathbb Z_p$, a temporal background sees only a $\mathbb Z_2$ part of the anomaly. It is nontrivial if and only if $p$ and $N$ contain the same power of $2$.
\end{itemize}

\paragraph{Geometry and flat sectors.}
Conventions follow Ref.~\cite{Anber:2026dmq} and are collected in the End Matter. The topological charge is $Q=\frac{1}{8\pi^2}\int\tr F\wedge F$, $\tau_a$ are the color Pauli matrices, and $\sigma_{x,y,z}$ are the left-invariant one-forms on $S^3$.

Write a point of $\mathbb R^4$ as $x=|x|\,\hat x$, with $\hat x$ on the angular $S^3\cong SU(2)$. Identifying $\mathbb R^4$ with the quaternions, $X\equiv x^0I_2+ix^i\tau_i$, we have $X^\dagger X=|x|^2I_2$, so $X=|x|\,g$ with $g\in SU(2)$. In Euler angles $(\theta,\varphi,\psi)$, with $\psi\in[0,4\pi)$, one has $g^{-1}dg=i\,\tau_a\sigma_a$ [Eq.~\eqref{eq:g}]. The flat metric is $ds^2=d|x|^2+|x|^2(\sigma_x^2+\sigma_y^2+\sigma_z^2)$. The reflection $x\to-x$ is $X\to-X$, which acts on each angular sphere as the antipodal map $\psi\to\psi+2\pi$, under which $g\to-g$. The quotient $S^3/\mathbb Z_2=\RP$ is described by the same angles with $0\le\psi<2\pi$. The loop $c$ that runs $\psi:0\to2\pi$ at fixed $(\theta,\varphi)$ joins $\hat x$ to $-\hat x$. It is open on $S^3$, closed on $\RP$, and generates $\pi_1(\RP)=\mathbb Z_2$.

We now describe the classical vacua of the $SU(2)$ theory on $\RP$. Three steps are involved.

\emph{(1) One bundle, two flat connections.} Because $SU(2)\cong S^3$ has homotopy groups $\pi_0=\pi_1=\pi_2=0$, every $SU(2)$ color bundle over a three-manifold is trivial \cite{Freed:1992vw}. On $\RP$ the gauge field is therefore a single, globally defined $\mathfrak{su}(2)$-valued one-form, with no patching and no 't~Hooft flux. The vacua are distinguished not by the bundle but by the flat connections on it \cite{Freed:1992vw}. A flat connection is fixed, up to gauge, by its holonomy $h_c$ around the generator $c$ of $\pi_1(\RP)=\mathbb Z_2$. Since $c^2=1$, $h_c^2=I$. In $SU(2)$ the eigenvalues of $h_c$ are then equal and $\pm1$, so $h_c=\pm I$. These two central elements are not conjugate, so they label two distinct sectors:
\begin{equation}
\ell=0:\ A_0=0,\qquad \ell=1:\ A_1=-i\,g\,dg^{-1} .
\label{eq:sectors}
\end{equation}

\emph{(2) Why $A_1$ is a nontrivial vacuum.} On the covering $S^3$, $A_1$ is the gauge transform of $A_0=0$ by $g$, so it is flat. Under the antipodal map $g\to-g$, and the sign cancels in $g\,dg^{-1}$, so $A_1$ is a well-defined flat connection on $\RP$. It is nevertheless not gauge equivalent to $A_0$ there: since $g(-x)=-g(x)$, $g$ is not a single-valued map $\RP\to SU(2)$, hence not a gauge transformation of the $SU(2)$ theory. This is detected gauge invariantly by the holonomy: parallel transport along $c$, which runs from $x$ to $-x$, gives $h_c=g(\psi+2\pi)\,g(\psi)^{-1}=-I$.

\emph{(3) The Chern--Simons invariant.} A direct integration (End Matter) gives
\begin{equation}
\CS(A_1)-\CS(A_0)=-\frac{1}{24\pi^2}\int_{\RP}\tr(g^{-1}dg)^3=\frac12 .
\label{eq:CS}
\end{equation}
The integral over $S^3$ would be $1$, the degree of $g$. On $\RP$ only half of the sphere is covered. A genuine gauge transformation $h:\RP\to SU(2)$ shifts $\CS$ by $\frac{1}{24\pi^2}\int_{\RP}\tr(h^{-1}dh)^3$, which is the integer degree of $h$. Hence $\CS\bmod1$ is gauge invariant, and its value $0$ or $\tfrac12$ distinguishes the sectors. The fractional value is possible only because $g$ is not a gauge function.

This is the key difference from $\mathbb T^3$. There, the flat connections related by center transformations are Abelian, for instance $A=\frac{\pi}{L}\tau_3dx$ with holonomy $-I$ around one cycle. Such a connection has $dA=0$ and $A\wedge A\wedge A=0$, hence $\CS=0$, so a center transformation does not change $\CS$ and the anomaly is invisible. A relative $\CS$ of $\tfrac12$ appears only after a spatial twist is imposed~\cite{tHooft:1979rtg,Cox:2021vsa}. On $\RP$ the torsion of $H_1(\RP)=\mathbb Z_2$ supplies it. The cycle $c$ has order two, and its self-linking number $\mathrm{lk}(c,c)=\tfrac12$ is exactly the fractional $\CS$ of a flat connection with holonomy $-I$ around it. 
\paragraph{The center operator.}
The center operator is a map from Hilbert space to itself. Its building block is the function $g$ of Eq.~\eqref{eq:g}, which, as emphasized above, is not an $SU(2)$-single-valued function on $\RP$. Its adjoint action, however, is single-valued, because $\mathrm{Ad}(-g)=\mathrm{Ad}(g)$. Define
\begin{equation}
U:\ A\mapsto gAg^{-1}-i\,g\,dg^{-1},\qquad \Phi_{\rm adj}\mapsto g\,\Phi_{\rm adj}\,g^{-1}.
\label{eq:U}
\end{equation}
Because the sign of $g$ cancels in both expressions, $U$ maps $SU(2)$ configurations on $\RP$ to $SU(2)$ configurations on $\RP$. It preserves the action and commutes with the Hamiltonian, since every field is in the adjoint. Because $g$ is not an allowed gauge transformation, $U$ is a genuine symmetry rather than a redundancy. It is the $\mathbb Z_2$ center ($1$-form) symmetry operator supported on $\mathbb{RP}^2$, the Poincar\'e dual of $c$ \cite{Freed:2006ya,Freed:2006yc, Gaiotto:2014kfa}. Up to gauge transformations, $g$ can be deformed to jump by $-I$ only across $\mathbb{RP}^2$, like 't~Hooft's electric-flux operator on a two-torus of the untwisted $\mathbb T^3$~\cite{tHooft:1979rtg}. Gauge invariantly, local operators and adjoint Wilson loops are invariant, while the holonomy transforms as $h_c\mapsto-g\,h_cg^{-1}$, so the fundamental Wilson loop $W_c=\tfrac12\tr\,h_c$, which crosses $\mathbb{RP}^2$ once, obeys $UW_cU^{-1}=-W_c$. This is the defining action of a $\mathbb Z_2$ $1$-form symmetry. Semiclassically, $W_c=(-1)^\ell$ on sector $\ell$, and $U$ exchanges the two values. Gauging $U$ would give the $SO(3)$ theory and identify the two sectors. In the $SU(2)$ theory, $U$ maps $A_0$ to $A_1$ and shifts the CS functional by a constant (End Matter):
\begin{equation}
U\,\CS\,U^{-1}=\CS+\tfrac12 .
\label{eq:UCS}
\end{equation}
A nontrivial $SO(3)$ bundle appears only in the Euclidean trace $\mathrm{Tr}[U\cdots]$. Gluing the ends of $\RP\times[0,\beta]$ with $U$ produces the mapping torus $\RP\times_US^1$. On this four-manifold the transition function is $SO(3)$-valued, so the bundle has $w_2\neq0$, with a leg only along the time direction, and charge $Q=\tfrac12$. Every spatial slice still carries the trivial $SU(2)$ bundle.

\paragraph{Anomalies and exact consequences.}
\emph{$SU(2)$ SYM theory.} With one adjoint Weyl fermion, a charge-$Q$ configuration carries $4Q$ zero modes, and the conserved $\mathbb Z_4^\chi$ generator is $X=\exp[\frac{i\pi}{2}(Q_5-4\,\CS)]$, where $Q_5$ is the gaugino number \cite{Cox:2021vsa,Anber:2023mlc}. Since $U$ commutes with $Q_5$, Eq.~\eqref{eq:UCS} gives
\begin{equation}
XUX^{-1}U^{-1}=-1 .
\label{eq:anomaly}
\end{equation}

\emph{YM theory at $\theta$.} Physical states are invariant under small gauge transformations (Gauss's law) and, since large gauge transformations commute with all gauge-invariant operators, transform as $\Psi[A^h]=e^{i\theta W(h)}\Psi[A]$ under those of degree $W(h)\in\mathbb Z$; $\theta$ labels the superselection sector, whose lowest states are the $\theta$-vacua. Time reversal $\mathsf T$ maps $\theta\to-\theta$, and the operator $e^{2\pi i\,\CS}$, which is invariant under large gauge transformations, shifts $\theta$ by $2\pi$. At $\theta=\pi$ the combination $\mathsf T'=e^{2\pi i\CS}\mathsf T$ is a symmetry~\cite{Gaiotto:2017yup,Cox:2021vsa}. Since $U$ commutes with $\mathsf T$,\footnote{If $\theta$ is imposed through $\Psi[A^h]=e^{i\theta W(h)}\Psi[A]$, the geometric map $(U\Psi)[A]=\Psi[A^g]$ obeys $U^2=e^{i\theta}$, since $g^2$ is a degree-one gauge transformation on $\RP$; at $\theta=\pi$ the time-reversal symmetry is $\mathsf T$, and $e^{-i\theta/2}U$ squares to one and anticommutes with it. After the unitary rotation $\Psi\to e^{-i\theta\,\CS}\Psi$, states are invariant under all gauge transformations and $\theta$ appears in $H$; then $U^2=1$, $U$ commutes with $\mathsf T$, and the symmetry at $\theta=\pi$ is $\mathsf T'=e^{2\pi i\CS}\mathsf T$. We work in the latter frame. The degeneracy and $\mathcal Z_U=0$ are the same in both.} Eq.~\eqref{eq:UCS} gives
\begin{equation}
U\mathsf T'U^{-1}\mathsf T'^{-1}=e^{2\pi i\cdot\frac12}=-1 .
\label{eq:anomalyYM}
\end{equation}

Equations~\eqref{eq:anomaly} and~\eqref{eq:anomalyYM} are the mixed chiral--center and $\mathsf T$--center anomalies, realized as operator algebras on the untwisted $SU(2)$ Hilbert space of $\RP$. They hold at every radius $R$ and have two exact consequences.

(a) \emph{Degeneracy.} $X$ (or $\mathsf T'$) maps $U=+1$ eigenstates to $U=-1$ eigenstates of the same energy. Every level of SYM theory, and of YM theory at $\theta=\pi$, is at least doubly degenerate on $\RP$ of any size.

(b) \emph{Vanishing lens-space partition function.} For $N=2$, $\RP=L(2,1)$ and the twisted partition function of Ref.~\cite{Hayashi:2026brp} is
\begin{equation}
\mathcal Z_U(\beta)\equiv\frac{\mathrm{Tr}\,[U e^{-\beta H}]}{\mathrm{Tr}\,e^{-\beta H}} .
\end{equation}
In SYM theory, cyclicity and Eq.~\eqref{eq:anomaly} give $\mathrm{Tr}[Ue^{-\beta H}]=\mathrm{Tr}[XUX^{-1}e^{-\beta H}]=-\mathrm{Tr}[Ue^{-\beta H}]$, and likewise with $(-1)^F$ inserted. In YM theory at $\theta=\pi$, Eq.~\eqref{eq:anomalyYM} with the antiunitarity of $\mathsf T'$ makes the trace purely imaginary, while $U^2=1$ and $[U,H]=0$ make it real (End Matter). In both cases (see \cite{Christensen:2025ktc} for a general statement)
\begin{equation}
\mathcal Z_U(\beta)=0\qquad\text{for all }R,\beta .
\label{eq:ZU0}
\end{equation}
Ref.~\cite{Hayashi:2026brp} associates $\mathcal Z_U\to\text{const}\neq0$ at large volume with monopole condensation; Eq.~\eqref{eq:ZU0} does not contradict this. The two confining vacua carry $\mathbb Z_2^{(1)}$ SPT levels $k=0,1$, with $\langle U\rangle_k=e^{i\pi k\int\frac12\mathcal P(B)}=(-1)^k$ since $\int\frac12\mathcal P(B)$ is odd on $\RP\times_US^1$: each is maximally condensed, and the two cancel in the trace.

\paragraph{Small $\RP$: tunneling and $\mathcal Z_U$ in closed form.}
At $R\Lambda\ll1$, where $\Lambda$ is the strong scale, the coupling is weak. Because $\mathrm{Ad}(-I)=1$, the fluctuation spectra around $A_0$ and $A_1$ are identical, so the two perturbative ground states $|0\rangle,|1\rangle$ are degenerate to all orders. Both holonomies are central, so $SU(2)$ is unbroken and there are no light $W$ bosons; since $H^1(\RP;\mathbb R)=0$ there are no flat directions either, and every gluon mode is gapped at $\sim 1/R$. The degeneracy is lifted by tunneling. Writing the flat metric as $ds^2=d|x|^2+|x|^2(\sigma_x^2+\sigma_y^2+\sigma_z^2)$ and setting $|x|=Re^{\tau/R}$ gives $ds^2=e^{2\tau/R}\big[d\tau^2+R^2(\sigma_x^2+\sigma_y^2+\sigma_z^2)\big]$. Thus $\mathbb R^4\setminus\{0\}$ is conformally equivalent to the cylinder $\mathbb R\times S^3_R$, with the origin at $\tau\to-\infty$. The Hodge star on two-forms is Weyl invariant, so self-dual solutions remain self-dual, with the same action and charge. Since $x\to-x$ preserves $|x|$, it leaves $\tau$ fixed and acts on each slice as the antipodal map. Instantons on the cylinder over $\RP$ are therefore the $\mathbb Z_2$-invariant instantons on $\mathbb R^4\setminus\{0\}$.

Such an instanton exists. The self-dual BPST instanton centered at the origin is, in regular gauge,
\begin{equation}
A=\frac{x^2}{x^2+\rho^2}\,A_1\,,
\label{eq:BPST}
\end{equation}
where $\rho \in (0, \infty)$ is the size modulus. 
It is smooth at the origin, tends to $A_1$ on $S^3_\infty$, and carries $Q=1$ on $\mathbb R^4$. It is invariant under $x\to-x$ with no gauge transformation, because the reflection acts on it as a central, hence trivial, constant gauge rotation (End Matter). Its four centered moduli, the size and three $SO(3)\cong\RP$ orientations, form the cone $\mathbb R^4/\mathbb Z_2$ itself.

On $\mathbb R\times S^3_R$, Eq.~\eqref{eq:BPST} is the familiar $Q=1$ tunneling between $A_0$ and its large gauge transform $A_1$. On $\RP$, $A_1$ is not pure gauge, and the same solution becomes a genuine transition between the sectors $\ell=0,1$, with half the charge. Rewritten in terms of $\tau$, the one-form~\eqref{eq:BPST} reads
\begin{equation}
A(\tau)=\frac{A_1}{1+e^{-2(\tau-\tau_0)/R}} ,
\end{equation}
a self-dual tunneling event on $\mathbb R\times\RP_R$ from $A_0$ to $A_1$ [Fig.~\ref{fig:completions}(b)], with $Q=\tfrac12$ and action $4\pi^2/g^2$ (a direct check of self-duality is given in the End Matter). Unlike the fractional instantons of twisted $\mathbb T^3\times\mathbb R$, this event is known in closed form\footnote{See, e.g., \cite{GarciaPerez:1992fj,DasilvaGolan:2022jlm} for the lattice construction of fractional instantons on the twisted $\mathbb T^3\times\mathbb R$.}.  The instanton size $\rho$ appears as the tunneling time $\tau_0=R\ln(\rho/R)$. Each event flips the sector. Applying $U$ to this solution gives a self-dual event from $A_1$ to $A_0$, since $U$ exchanges the sectors and preserves the action. Reversing $\tau$ turns both into anti-self-dual events with $Q=-\tfrac12$. In the Euclidean amplitude $\langle\ell'|e^{-TH}|\ell\rangle$, each event therefore carries the weight $e^{-4\pi^2/g^2}e^{\pm i\theta/2}$, times a one-loop determinant and the measure over its four moduli. Integrating the tunneling time $\tau_0$ over $[0,T]$ gives a factor $T$, and the three orientations together with the determinant give a constant $K$ of dimension $1/R$. On the two-dimensional space spanned by $|0\rangle$ and $|1\rangle$, define the phase of $|1\rangle$ by $|1\rangle\equiv U|0\rangle$; then $U|1\rangle=U^2|0\rangle=|0\rangle$. Since $[U,H]=0$, the flip amplitudes $0\to1$ and $1\to0$ are equal. A single flip therefore acts on this space as $U$ itself, i.e., as the matrix $\big(\begin{smallmatrix}0&1\\1&0\end{smallmatrix}\big)$. Summing over a dilute gas of $n$ events gives
\begin{eqnarray}\nonumber
\langle\ell'|e^{-TH}|\ell\rangle&=&e^{-E_0T}\sum_n\frac{(2KT\cos\frac\theta2)^n}{n!}\,\langle\ell'|U^n|\ell\rangle\\
&=&e^{-E_0T}\langle\ell'|e^{2KT\cos\frac\theta2\,U}|\ell\rangle,
\end{eqnarray}
where $E_0$ is the common perturbative energy. Hence
\begin{equation}
H_{\rm eff}=E_0-2K\cos\tfrac\theta2\,U,\qquad K\propto\frac{1}{R}\,\frac{(R\Lambda)^{11/3}}{g^4(1/R)}.
\end{equation}
The power $(R\Lambda)^{11/3}=e^{-4\pi^2/g^2(1/R)}$ follows from one-loop running with $b_0=22/3$, and $g^{-4}$ comes from the four bosonic collective coordinates. The gas is dilute because $KR\ll1$ for $R\Lambda\ll1$. The energy eigenstates are the $U$-eigenstates $|k\rangle=\big(|0\rangle+(-1)^k|1\rangle\big)/\sqrt2$, with $U|k\rangle=(-1)^k|k\rangle$ and $E_k=E_0-(-1)^k2K\cos\frac\theta2$. Since $UW_cU^{-1}=-W_c$, the Wilson loop, diagonal on the sectors ($W_c=(-1)^\ell$), maps $|k=0\rangle$ to $|k=1\rangle$, so $\langle W_c(\tau)W_c(0)\rangle\propto e^{-(E_1-E_0)\tau}$ and $E_1-E_0=4K\cos\frac\theta2$ is the energy cost of one unit of $\mathbb Z_2$ electric flux along the torsion cycle $c$, the $\RP$ analogue of 't~Hooft's electric-flux energy on $\mathbb T^3$~\cite{tHooft:1979rtg}. In a confining phase one expects it to grow as $\sigma\pi R$ at large $R$, where $\pi R$ is the length of $c$; at small $R$ the flux is delocalized and its cost is set by tunneling. For $\beta\gg R$ only the states $|k=0,1\rangle$ contribute, so $\mathrm{Tr}[Ue^{-\beta H}]=\sum_k(-1)^ke^{-\beta E_k}\propto\sinh(2\beta K\cos\frac\theta2)$ and $\mathrm{Tr}\,e^{-\beta H}\propto\cosh(2\beta K\cos\frac\theta2)$, giving\footnote{On $\mathbb R^3\times S^1$, monopole instantons also predict a twofold degeneracy at $\theta=\pi$ \cite{Unsal:2012zj}.} 
\begin{equation}
\mathcal Z_U(\beta)=\tanh\!\Big(2\beta K\cos\frac\theta2\Big).
\label{eq:tanh}
\end{equation}
It vanishes at $\theta=\pi$, as Eq.~\eqref{eq:ZU0} requires, and the levels cross there. Under $\theta\to\theta+2\pi$, the labels $k=0,1$ are exchanged. This is the $\RP$ counterpart of 't~Hooft's electric-flux energies on twisted $\mathbb T^3$ \cite{vanBaal:thesis}, obtained here without a twist. 
\paragraph{A semiclassical cap: EH space.}
In SYM theory, the same $Q=\tfrac12$ event carries two gaugino zero modes. It therefore cannot contribute to $H$,  but it gives the matrix element of $\cO=\tr(\lambda\lambda)$. On the two lowest states, $\cO$ has $X$-charge $-1$, so it is off-diagonal in $\ell$, and it is local, so it commutes with $U$. Hence $\langle1|\cO|0\rangle=\langle0|\cO|1\rangle\equiv C$ and
\begin{equation}
\cO_{\rm eff}=C\,U,\qquad \langle k|\cO|k\rangle=(-1)^kC,
\label{eq:Oeff}
\end{equation}
with $C\propto e^{-4\pi^2/g^2+i\theta/2}$. The cylinder considered in the YM case cannot, however, fix $C$. On the round $\mathbb R\times\RP_R$, supersymmetry requires a $U(1)_R$ background that pure SYM lacks, so $C$ is unprotected and depends on $R$. What is needed is a geometry that keeps the $Q=\tfrac12$ saddle with its two zero modes, preserves supersymmetry without an $R$-symmetry, and is asymptotically flat, so that $C$ becomes the flat-space condensate.

Such a geometry is obtained by completing the cone $\mathbb R^4/\mathbb Z_2$ differently (Fig.~\ref{fig:completions}). The cylinder pushes the tip to $\tau\to-\infty$ and keeps $\RP$ at fixed size. Resolving the tip instead replaces it by a smooth bolt and keeps the asymptotics flat. These are two fixed backgrounds carrying the same $\mathbb Z_2$-invariant BPST configuration, with $Q=\CS(A_1)-\CS(\text{interior})=\tfrac12$ in both; they are not contributions to be summed\footnote{The partition function is obtained as the path-integral over a fixed 4-D geometry; we do not sum over geometries in ordinary QFT.}. In YM theory, the resolved geometry is not useful quantitatively: nothing protects the amplitude and the size integral diverges in the infrared, whereas on the cylinder the radius provides the cutoff.

On the cone, before resolution, the instanton~\eqref{eq:BPST} has $Q=\tfrac12$, asymptotic holonomy $-I$, and centered moduli space $\mathbb R^4/\mathbb Z_2$, as found above. What is new in SYM theory are the fermion zero modes. Two of the four undotted gaugino zero modes survive. (Dotted zero modes vanish in a self-dual background~\cite{Anber:2026dmq}.) The lift $(L,R)=(-I_2,I_2)$ preserves the supercharges $\bar Q_{\dot\alpha}$ and acts as $-1$ on undotted spinors. The supersymmetric modes $\lambda_\alpha\propto F_\alpha{}^\beta\xi_\beta$, with $\xi$ constant, are even and projected out. The superconformal modes $\lambda_\alpha\propto F_\alpha{}^\beta\epsilon_\beta(x)$, with the undotted conformal Killing spinor $\epsilon_\beta(x)=x_a\hat\sigma^a_{\beta\dot\beta}\bar\epsilon^{\dot\beta}$, are odd and survive. The cone is scale invariant, so they are generated by broken superconformal symmetries (End Matter).

The minimal smooth resolution of this cone that preserves $\bar Q_{\dot\alpha}$ is the hyperk\"ahler Eguchi--Hanson (EH) space, with metric
\begin{equation}
ds^2=f^2dr^2+r^2\big[\sigma_x^2+\sigma_y^2+f^{-2}\sigma_z^2\big],\qquad f^{-2}=1-\frac{a^4}{r^4},
\end{equation}
with $r\ge a$, $\psi\in[0,2\pi)$, and a bolt $S^2$ at $r=a$. It supports the $SU(2)$ self-dual configuration
\begin{equation}
A=\tau_3\,\frac{a^2}{r^2}\,\sigma_z,\qquad Q(r)=\frac12\Big(1-\frac{a^4}{r^4}\Big),
\label{eq:EHsol}
\end{equation}
the $t=a$ endpoint of a family with size modulus $t\in[a,\infty)$~\cite{Anber:2026dmq}. Here $Q(r)$ is the charge enclosed within radius $r$. In the gauge regular at the bolt, the Wilson loop around the $\psi$-circle, $W(r)=\exp[-i\pi\tau_3(1-a^2/r^2)]$, runs from $I$ at the bolt to $-I$ at infinity. By Stokes' theorem, EH is a \emph{cap} that prepares a state on $\RP$ [Fig.~\ref{fig:completions}(c)] \cite{Anber:2026pfg}: its inner slice degenerates to the bolt, which supplies $\CS=0$, so $Q=\CS(A_1)-0=\tfrac12$. The moduli space is a copy of EH itself~\cite{Kronheimer:1990,Bianchi:1995xd,Bianchi:1996zj}: at $t=a$ the orientation orbit collapses to $SO(3)/U(1)=S^2$, the bolt. The parameter $a$ breaks scale invariance, and the two zero modes are now protected by the Atiyah--Patodi--Singer (APS) index~\cite{Atiyah:1975jf}, which equals two for every $a$ (End Matter).

\paragraph{The gaugino condensate.}
\emph{What EH computes.} With fixed data at infinity, the EH path integral prepares a state $|\Psi_\EH\rangle\in\Hil(\RP)$ on the asymptotic $\RP$. Inserting $\cO=\tr(\lambda\lambda)\equiv\tr(\epsilon^{\alpha\beta}\lambda_\beta\lambda_\alpha)$ gives $|\Psi_\EH[\cO]\rangle$. A semiclassical computation fixes the asymptotic holonomy, so it evaluates the sector components $Z^{(\ell)}[\,\cdot\,]=\langle\ell|\Psi_\EH[\,\cdot\,]\rangle$. Since the bolt carries $\CS=0$, a saddle of charge $Q$ ends in the sector with $\CS\equiv Q$: saddles with $Q\in\mathbb Z$ end in $\ell=0$, and saddles with $Q\in\tfrac12+\mathbb Z$ end in $\ell=1$. A saddle of charge $Q$ carries $4Q$ gaugino zero modes. The trivial saddle has none and contributes to $Z^{(0)}[1]$. The $Q=	\frac12$ saddle has two, saturated by $\cO$, and contributes to $Z^{(1)}[\cO]$. $X$-charge conservation makes this exact,
\begin{equation}
Z^{(0)}[\cO]=0=Z^{(1)}[1],\qquad Z^{(1)}[\cO]\neq0\neq Z^{(0)}[1],
\end{equation}
which is the EH counterpart of $\cO$ being off-diagonal in $\ell$, Eq.~\eqref{eq:Oeff}.

\emph{From sectors to vacua.} The condensate is therefore not defined sector by sector, $\langle\ell|\cO|\ell\rangle=0$; it needs a bra that overlaps both sectors, namely a vacuum. The Witten index of $SU(2)$ SYM theory is two~\cite{Witten:1982df}, and by Eq.~\eqref{eq:anomaly} $X$ and $U$ cannot be diagonalized simultaneously on the two vacua. In the confining phase, the clustering vacua are the $U$ eigenstates, $U|k\rangle=(-1)^k|k\rangle$ (the standard picture, not derived here). Since $|1\rangle=U|0\rangle$, their components obey $\langle k|1\rangle=(-1)^k\langle k|0\rangle$, and
\begin{equation}
\langle\cO\rangle_k=\frac{\langle k|\Psi_\EH[\cO]\rangle}{\langle k|\Psi_\EH[1]\rangle}=(-1)^k\,\frac{Z^{(1)}[\cO]}{Z^{(0)}[1]} .
\label{eq:condk}
\end{equation}
The anomaly fixes the relative sign: since $XUX^{-1}=-U$, $X$ maps $|k\rangle$ to $|k+1\rangle$, and since $X\cO X^{-1}=-\cO$, $\langle\cO\rangle_{k+1}=-\langle\cO\rangle_k$. The magnitude is the ratio of the $Q=\tfrac12$ saddle, with its zero modes saturated by $\cO$, to the trivial saddle. It is the EH counterpart of the constant $C$, now computed on a supersymmetric background.

\emph{From EH to flat space.} Two Ward identities make Eq.~\eqref{eq:condk} a flat-space vacuum quantity. First, $\partial_\mu\cO$ is $\bar Q$-exact, so $\cO$ can be inserted far from the bolt, where the geometry is flat. Second, the metric derivative $\partial_ag_{\mu\nu}$ is a normalizable transverse traceless mode whose coupling to the stress tensor $T_{\mu\nu}$ is $\bar Q$-exact, so the result does not depend on the bolt size $a$~\cite{Anber:2026dmq} (End Matter). The EH condensate was first studied in Ref.~\cite{Bianchi:1995ad}. Its full collective-coordinate evaluation with Pauli--Villars regularization~\cite{Anber:2026dmq} gives
\begin{equation}
\langle\tr(\lambda\lambda)\rangle_k=16\pi^2\Lambda_k^3,\qquad \Lambda_k^3\equiv\frac{\mu^3}{g^2}e^{-\frac{4\pi^2}{g^2}+\frac i2(\theta+2\pi k)}.
\label{eq:result}
\end{equation}
 This is precisely the weak-coupling instanton (WCI) value. On $\mathbb R^4$ the BPST instanton has four zero modes and cannot saturate $\cO$, so the strong-coupling instanton (SCI) route extracts $\langle\cO\rangle$ from $\langle\cO\cO\rangle$ by clustering, and obtains a value smaller by $\sqrt{4/5}$~\cite{Amati:1988ft,Finnell:1995dr,Hollowood:1999qn}. The single EH saddle needs no clustering and sides with WCI, consistent with Refs.~\cite{Hollowood:1999qn,Davies:1999uw,Anber:2022qsz,Anber:2024mco,Bonanno:2024bqg,Hayashi:2026srq}.

\paragraph{Generalizations.}
The construction extends to $SU(N)$ on lens space $L(p,1)\equiv S^3/\mathbb Z_p$, where $\mathbb Z_p$ acts on $\mathbb C^2\equiv \mathbb R^4$ by $(z_1,z_2)\to\zeta(z_1,z_2)$, $\zeta=e^{2\pi i/p}$, i.e.\ $\psi\to\psi+4\pi/p$. A center transformation along $c$ must obey $z^p=1$, so only the subgroup $\mathbb Z_d$ acts, with $d=\gcd(p,N)$. The flat sectors are $h_c=\mathrm{diag}(\zeta^{m_1},\dots,\zeta^{m_N})$, with $\CS\equiv-\sum_im_i^2/2p$ for a lift with $\sum_im_i=0$~\cite{KirkKlassen,Razamat:2013opa,Duan:2024xbb}. The generator of $\mathbb Z_d$ shifts every $m_i$ by $p/d$. After re-lifting, this shifts $\CS$ uniformly by
\begin{equation}
\Delta\CS\equiv-\frac{pN(N+1)}{2d^2}\pmod 1 .
\label{eq:dCSgen}
\end{equation}
Since $d\mid p$ and $d\mid N$, $\Delta\CS\in\frac12\mathbb Z$. A purely temporal background, $B=a\cup dt$, has $\int B\cup B\equiv0\pmod N$; only the Pontryagin-square refinement survives, through the torsion of $L(p,1)$, so it detects at most a $\mathbb Z_2$ part of the mixed anomaly. The phase $XUX^{-1}U^{-1}=U\mathsf T'U^{-1}\mathsf T'^{-1}=e^{2\pi i\Delta\CS}$ is $-1$ if and only if
\begin{equation}
v_2(p)=v_2(N)\ge1,
\label{eq:v2}
\end{equation}
where $v_2(n)$ is the exponent of the largest power of $2$ dividing $n$. When Eq.~\eqref{eq:v2} does not hold, the lens space is anomaly blind. On $\RP$ the criterion gives $N\equiv2\pmod4$.

Torsion need not come from curved space. For $SU(2)$, the anomaly is visible on a closed three-manifold $M$ whenever some $a\in H^1(M;\mathbb Z_2)$ has $\int_Ma^3\neq0$ \cite{Turaev:1984}. Among flat manifolds  \cite{Conway:2003tp}, we find that this holds for the quarter-turn space, a cubic lattice that is periodic in $x$ and $y$ and closed in $z$ after a $90^\circ$ rotation, but not for $\mathbb T^3$, the half-turn space, or the Hantzsche--Wendt space. We therefore expect that Hamiltonian lattice gauge theory, or a quantum simulator, with this screw boundary condition would exhibit the exact twofold degeneracies of $SU(2)$ SYM theory and of YM theory at $\theta=\pi$ without any 't~Hooft flux, with the other flat manifolds as built-in controls.

\begin{acknowledgments}
We would like to thank I\~naki Garc\'{\i}a Etxebarria, Erich Poppitz, and  Yuya Tanizaki for discussions and valuable comments on the manuscript. Also, we would like to thank Erich Poppitz for the hospitality at the University of Toronto, where part of this project was conducted. 
Opus 5.5 Claude by Anthropic and Astra-6 ChatGPT by OpenAI were used for calculation checks and edits. This work is supported by the STFC consolidated grant UKRI4226.
\end{acknowledgments}


\clearpage
\section*{End Matter}

\paragraph{Appendix A: Conventions and the Chern--Simons invariant on $\RP$.}
We follow Ref.~\cite{Anber:2026dmq}: Hermitian gauge fields with $F=dA+iA\wedge A$, color Pauli matrices $\tau_a$, spin matrices $\hat\sigma^a=(I_2,i\vec\sigma)$, and topological charge $Q=\frac{1}{8\pi^2}\int\tr F\wedge F$. The left-invariant one-forms on $S^3$ are
\begin{equation}
\begin{aligned}
\sigma_x&=\tfrac12\left(\sin\psi\,d\theta-\sin\theta\cos\psi\,d\varphi\right),\\
\sigma_y&=\tfrac12\left(-\cos\psi\,d\theta-\sin\theta\sin\psi\,d\varphi\right),\\
\sigma_z&=\tfrac12\left(d\psi+\cos\theta\,d\varphi\right),
\end{aligned}
\label{eq:LI}
\end{equation}
with $d\sigma_x=2\sigma_y\wedge\sigma_z$ (cyclic) and $\sigma_x\wedge\sigma_y\wedge\sigma_z=-\frac18\sin\theta\,d\theta\wedge d\varphi\wedge d\psi$, so that $\int_{\RP}\sigma_x\wedge\sigma_y\wedge\sigma_z=-\pi^2$. We parametrize
\begin{equation}
g=e^{i\varphi\tau_3/2}e^{i\theta\tau_2/2}e^{i(\psi+\pi)\tau_3/2},
\label{eq:g}
\end{equation}
with $0\le\theta\le\pi$, $0\le\varphi<2\pi$, $0\le\psi<4\pi$; a direct computation gives $g^{-1}dg=i\tau_a\sigma_a$, the shift $\psi\to\psi+\pi$ aligning the Maurer--Cartan form with Eq.~\eqref{eq:LI}. With $\CS(A)=\frac{1}{8\pi^2}\int\tr\big(A\wedge dA+\frac{2i}{3}A\wedge A\wedge A\big)$, so that $d\,\CS=\frac{1}{8\pi^2}\tr F\wedge F$, a pure gauge $-i\,h^{-1}dh$ has $\CS=\frac{1}{24\pi^2}\int\tr(h^{-1}dh)^3$. Using $\tr(\tau_a\tau_b\tau_c)=2i\epsilon_{abc}$, $\tr(g^{-1}dg)^3=12\,\sigma_x\wedge\sigma_y\wedge\sigma_z$. Since $A_1=-i\,g\,dg^{-1}$ corresponds to $h=g^{-1}$,
\begin{equation}
\CS(A_1)=-\frac{12}{24\pi^2}\int\sigma_x\wedge\sigma_y\wedge\sigma_z=
\begin{cases}1,& S^3:\ \psi\in[0,4\pi),\\ \tfrac12,& \RP:\ \psi\in[0,2\pi).\end{cases}
\end{equation}

\paragraph{Appendix B: center operator and the vanishing traces.}
Because $g(\psi+2\pi)=-g(\psi)$ and $\mathrm{Ad}(-g)=\mathrm{Ad}(g)$, $U$ maps antipodally invariant adjoint configurations to antipodally invariant ones. The holonomy transforms as $h\mapsto-ghg^{-1}$, and $U(A_0)=A_1$. On a closed three-manifold $\CS(A^U)=\CS(A)+\CS(A_1)$, which gives Eq.~\eqref{eq:UCS}. Then $UXU^{-1}=X\,e^{\frac{i\pi}{2}(-4)\frac12}=-X$ and $U\mathsf T'U^{-1}=e^{2\pi i(\CS+\frac12)}\mathsf T=-\mathsf T'$.

For the traces, note that $U^2=1$ and $[U,H]=0$, so in a basis that diagonalizes both, $\mathrm{Tr}[Ue^{-\beta H}]=\sum_n\pm e^{-\beta E_n}$ is real. In SYM theory, $X$ is unitary and commutes with $H$, so $\mathrm{Tr}[Ue^{-\beta H}]=\mathrm{Tr}[XUX^{-1}e^{-\beta H}]=-\mathrm{Tr}[Ue^{-\beta H}]=0$. In YM theory at $\theta=\pi$, $\mathsf T'$ is antiunitary and commutes with $H$, and $\mathsf T'U\mathsf T'^{-1}=-U$. For an antiunitary operator, $\mathrm{Tr}[\mathsf T'\mathcal A\,\mathsf T'^{-1}]=\mathrm{Tr}[\mathcal A]^*$, so $\mathrm{Tr}[Ue^{-\beta H}]^*=-\mathrm{Tr}[Ue^{-\beta H}]$. The trace is both real and imaginary, hence zero.

\paragraph{Appendix C: the tunneling solution.}
The instanton~\eqref{eq:BPST} survives the quotient because of its symmetries. Rotations act as $X\to LXR^\dagger$, $(L,R)\in SU(2)_L\times SU(2)_R$, i.e., $g\to LgR^\dagger$. Eq.~\eqref{eq:BPST} is exactly $SU(2)_R$ invariant, while an $SU(2)_L$ rotation equals the constant gauge rotation $A\to LAL^{-1}$, the familiar locking of the undotted spinor index to color. The element $L=-I_2$, with $R=I_2$, is the reflection $x\to-x$, and its gauge action is central and therefore trivial. Hence Eq.~\eqref{eq:BPST} is invariant under $x\to-x$ with no gauge transformation. The same fact makes the orientation moduli $SU(2)/\mathbb Z_2\cong SO(3)\cong\RP$, so the four centered moduli, the size and three $SO(3)$ orientations, form the cone $\mathbb R^4/\mathbb Z_2$ itself.

Since $A_1=-i\,g\,dg^{-1}$ is flat, $dA_1=-iA_1\wedge A_1$. For $A=f(\tau)A_1$ this gives
\begin{equation}
F=f'\,d\tau\wedge A_1+i(f^2-f)\,A_1\wedge A_1 .
\end{equation}
Write $A_1=\tau_a\theta^a$. Flatness implies $d\theta^a=\epsilon_{abc}\theta^b\wedge\theta^c$, so the $\theta^a$ are right-invariant forms with the normalization of Eq.~\eqref{eq:LI}, and the round metric on $S^3_R$ is $R^2\sum_a(\theta^a)^2$. Using $\tau_b\tau_c=\delta_{bc}+i\epsilon_{bca}\tau_a$,
\begin{equation}
F=\tau_a\Big[\frac{f'}{R}\,E^0\wedge E^a+\frac{2f(1-f)}{R^2}\,E^b\wedge E^c\Big],
\end{equation}
with $E^0=d\tau$, $E^a=R\theta^a$, and $(a,b,c)$ cyclic. The right-invariant forms obey $\theta^b=-(\mathrm{Ad}_g)_{ba}\sigma_a$, hence $\theta^1\wedge\theta^2\wedge\theta^3=-\sigma_x\wedge\sigma_y\wedge\sigma_z$. With $|x|=Re^{\tau/R}$, the orientation of Ref.~\cite{Anber:2026dmq}, $\mathrm{vol}\propto-dr\wedge\sigma_x\wedge\sigma_y\wedge\sigma_z$, is $\mathrm{vol}\propto+E^0\wedge E^1\wedge E^2\wedge E^3$, so $\star(E^0\wedge E^a)=E^b\wedge E^c$. Self-duality is then $f'=2f(1-f)/R$, solved by $f=[1+e^{-2(\tau-\tau_0)/R}]^{-1}$. The opposite sign gives the anti-self-dual event from $A_1$ to $A_0$. Equivalently, the configuration is the conformal image of Eq.~\eqref{eq:BPST} with $\tau_0=R\ln(\rho/R)$, since a Weyl rescaling preserves the Hodge star on two-forms. Its charge is $Q=\CS(A_1)-\CS(A_0)=\tfrac12$. We use right-invariant forms because they make the regular-gauge instanton~\eqref{eq:BPST} self-dual in this orientation; that instanton is smooth at the orbifold point and asymptotes to $A_1$. The left-invariant representative $-i\,g^{-1}dg=\tau_a\sigma_a$ of the same sector has $\CS=-\tfrac12$ and is related to $A_1$ by a degree-one large gauge transformation. In it, the self-dual event is the singular-gauge solution $\rho^2/(x^2+\rho^2)\,\tau_a\sigma_a$ from $\ell=1$ to $\ell=0$, the type to which the EH solution~\eqref{eq:EHsol} belongs. The two descriptions are related by $A\to g^{-1}Ag-i\,g^{-1}dg$, which maps $f\,\tau_a\theta^a$ to $(1-f)\,\tau_a\sigma_a$. On $S^3$ this is a large gauge transformation; on $\RP$ it is the center map $U$ (up to the allowed transformation $g^{-2}$), so the two events are related by symmetry and carry equal weight.

\paragraph{Appendix D: zero modes.}
For the adjoint representation of $SU(N)$ on EH with asymptotic holonomy $h$, the APS index is
\begin{equation}
\mathcal I_{\rm adj}=2NQ+\frac{-(N^2-1)+\mathrm{sgn}(\mathrm{Hol})}{8},
\end{equation}
where the second term combines the gravitational and $\eta$-invariant contributions. Here $\mathrm{sgn}(\mathrm{Hol})=\sum_{\rm weights}(\pm1)$ counts the adjoint components that are periodic ($+1$) or antiperiodic ($-1$) around $c$. For $SU(2)$ with central $h=-I$, all three components are periodic, so $\mathrm{sgn}(\mathrm{Hol})=3$ and $\mathcal I_{\rm adj}=4\cdot\tfrac12+0=2$, independent of $a$. As $a\to0$ the explicit EH modes~\cite{Anber:2026dmq} have the profile $rt^2/(r^2+t^2)^2$ of the superconformal modes, not the profile $\rho^2/(r^2+\rho^2)^2$ of the supersymmetric ones. On $\mathbb R^4$ the BPST field strength scales as $|F|\propto\rho^2/(r^2+\rho^2)^2$. The supersymmetric modes $\lambda=F\xi$, with $\xi$ constant, inherit this profile and are nonzero at the origin. The superconformal modes $\lambda=F\epsilon(x)$, with $\epsilon(x)=x_a\hat\sigma^a\bar\epsilon$ linear in $x$, carry an extra factor of $r$, $|\lambda|\propto r\rho^2/(r^2+\rho^2)^2$, and vanish at the origin; this factor of $x$ is also what makes them odd under $x\to-x$. With the size $t$ in place of $\rho$, the EH modes at $a\to0$ reproduce the second profile \cite{Anber:2026dmq}, confirming that the two surviving zero modes are the superconformal ones.

\paragraph{Appendix E: independence of $a$.}
In the coordinate $\varrho^4=r^4-a^4$, the EH metric is $ds^2=\frac{\varrho^2}{R^2}d\varrho^2+R^2(\sigma_x^2+\sigma_y^2)+\frac{\varrho^4}{R^2}\sigma_z^2$ with $R^4=\varrho^4+a^4$. At fixed $\varrho$, $\partial_aR=a^3/R^3$, and $h_{\mu\nu}=\partial_ag_{\mu\nu}$ is, in the orthonormal frame,
\begin{equation}
h_{ab}=\frac{2a^3}{R^4}\,\mathrm{diag}(-1,1,1,-1).
\end{equation}
We checked that $h$ is traceless and divergence-free, and that $\int\sqrt g\,h_{ab}h^{ab}<\infty$. For a symmetric transverse traceless $h$, the Ferrara--Zumino multiplet gives
\begin{equation}
\tfrac12h^{\mu\nu}T_{\mu\nu}=\{\bar Q,\mathcal V\}+\nabla_\nu(\cdots)^\nu .
\end{equation}
The surface term falls off as $a^3/r^4$ against an area $\propto r^3$, and the bra, the ket, and $\cO$ are $\bar Q$-closed. Hence $\partial_a\langle\cO\rangle=0$; the trace anomaly drops out because $h$ is traceless.  The details on the derivation of Ward identities are in Ref.~\cite{Anber:2026dmq}.

\end{document}